\documentclass[aps,prl,showpacs,twocolumn]{revtex4-1}

\usepackage{graphicx}
\usepackage{latexsym}

\newcommand{\Tilde}{\raise.17ex\hbox{$\scriptstyle\sim$}}
\begin{document}


\title{Effects of long-range links on metastable states in a dynamic interaction network}

\author{Suhan Ree}
\email{suhan@physics.utexas.edu}
\affiliation{Center for Complex Quantum Systems and Department of Physics, University of Texas at Austin, Austin, TX, 78712, USA}
\affiliation{School of Liberal Arts and Science, Kongju National University, Yesan-Up, Yesan-Gun, Chungnam, 340-702, South Korea}

\date{\today}

\begin{abstract}
We introduce a model for random-walking nodes on a periodic lattice, where
the dynamic interaction network is defined from local interactions
and $E$ randomly-added long-range links.
With periodic states for nodes and an interaction rule of repeated averaging, 
we numerically find two types of metastable states at low- and 
high-$E$ limits, respectively, along with consensus states.
If we apply this model to opinion dynamics,
metastable states can be interpreted as sustainable diversities in our societies, and
our result then implies that, while diversities decrease and eventually disappear with more
long-range connections,
another type of states of diversities can appear when networks are 
almost fully-connected.
\end{abstract}



\maketitle

Complex systems usually consist of interconnected parts, forming complex networks,
and typically exhibit complex and unpredictable behavior.
While the concept of complex networks can be applied to various research areas, 
physicists have been trying to
find general frameworks for the network theory\cite{Albert:2002,Boccaletti:2006,Dorogovtsev:2008},
and their effort can be summed up in three categories.
First, generating mechanisms of networks satisfying certain statistical properties
as in small-world (SW) and scale-free networks\cite{Watts:1998,Newman:1999,Barabasi:1999},
and statistical measures on their structures have been studied.
Second, if nodes have states, dynamic behavior of nodes {\em on} given networks has been studied:
for example, opinion dynamics\cite{Barahona:2002,Laguna:2003,Suchecki:2005,Toivonen:2009} and 
synchronization problems\cite{Hong:2002,Heylen:2006} on SW networks.
Finally, dynamic behavior of both nodes and links for ``coevolving" 
networks\cite{Holme:2006,Gonzalez:2006,Benczik:2009}, where nodes and links evolve,
influencing each other, has been studied,
especially when time scales for node and link dynamics are comparable.
Here we present a simple model of coevolving networks that can mimic 
mobile agents in finite spaces\cite{Manrubia:2001,Gonzalez:2006,Fujiwara:2011}.
With additional long-range links, it produces two types of metastable 
states\cite{Suchecki:2005,Toivonen:2009,Benczik:2009}, 
and we focus on their dynamic behavior by finding their mean lifetimes.

\begin{figure}
\includegraphics[scale=0.8]{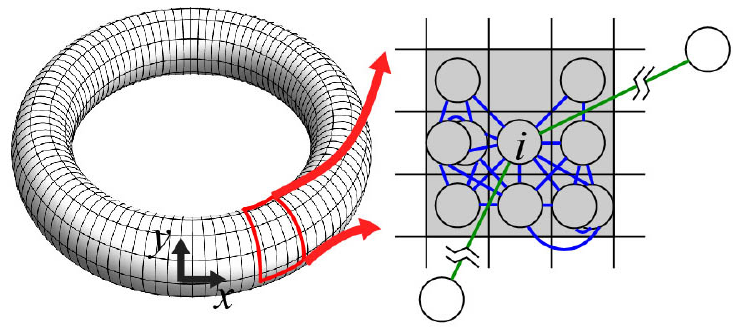}
\caption{\label{description}
(color online).
An illustration of a 2D periodic lattice.
Neighbors of node $i$ are in the shaded area, and
blue lines represent dynamic local interactions, while
green lines represent randomly-added long-range links.
}
\end{figure}

For our model, we start with a network defined in Ref.\ \onlinecite{Ree:2011} with minor modifications.
It considers $N$ nodes residing
on a two-dimensional (2D) periodic lattice ($X\times Y$) (see Fig.\ \ref{description}).
Each node $i$ ($1\le i\le N$) has 
its location ($x_i,y_i$), where $0\le x_i<X$ and $0\le y_i<Y$.
At each time step $t$ ($=0,1,2,\ldots$), all nodes move independently using 2D random walks,
and interact only with neighbors.
We call this network as a {\em dynamic interaction network}, and it is 
determined by (i) how neighbors are defined, and (ii) how nodes move on the lattice.
Here we use the Moore neighborhood; hence a node is a neighbor of node $i$ if that node
resides in any of the surrounding 8 locations or the current location of $i$ (see the shaded
area in Fig.\ \ref{description}).
And nodes move randomly and independently in four directions (east, west, south, and north) 
with equal probabilities.
This network is dynamic, and we will restrict ourselves to the case with $X=100$ and $Y=10$,
and $N=10^3(=XY)$, guaranteeing a giant component of connected nodes.

To this network, we add long-range bidirectional links (or {\em shortcuts}).
At $t=0$, $E$ time-invariant links are added randomly, meaning that every possible pair
of nodes has the equal probability of getting an additional link.
The number of all possible pairs of nodes is $N(N-1)/2$ $(\equiv E_{max})$,
and we use a normalized value $e=E/E_{max}$ ($0\le e \le 1)$ as a control parameter.
The network has two types of links: one representing short-range dynamic interactions, and 
the other representing randomly-chosen long-range static interactions.
When both types of links exist for any pair of nodes, they are regarded as one link.
We can use the $N\times N$ adjacency matrix ${\bf K}(t)$ to represent this network, 
where $K_{ij}(t)$ is 1 when there 
exists a link between $i$ and $j$ at time $t$, and 0 otherwise.

\begin{figure}
\includegraphics[scale=0.36]{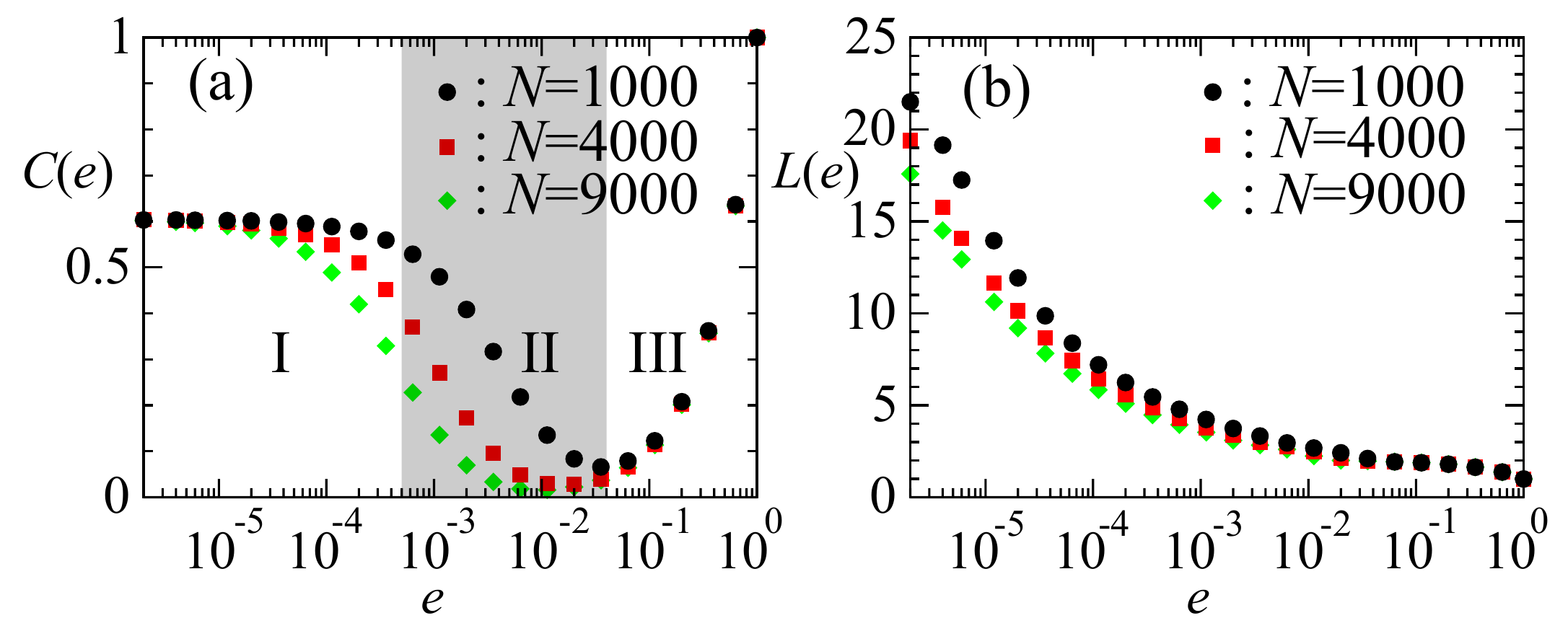}
\caption{\label{smallworld}
(color online).
For $N$ uniformly distributed nodes
when $(N,X,Y)$ is $(1000, 100, 10)$, $(4000,200,20)$, and $(9000,300,30)$,
we observe (a) $C(e)$ and (b) $L(e)$ versus $e$.
In (a), we divide $e$ into three regions for the case of $N=10^3$.
}
\end{figure}

When $e=0$, the network has only local interactions, 
and, when $e=1$, the network is fully-connected.
We can look at some of the basic network properties of this network: the average degree $k(e)$,
the clustering coefficient $C(e)$, and the average shortest path length $L(e)$,
when all nodes are uniformly distributed on the lattice\footnote{Because the motion of nodes 
is diffusive, the density of nodes on the lattice 
converges to the uniform distribution regardless of the initial positions of nodes.}.
We numerically find these values by averaging over $10^3$ realizations:
for $e=0$, $k(0)\simeq8.99$, $C(0)\simeq0.604$, and $L(0)\simeq25.8$.
The probability of having more than one component is noticeable when $e$ is close to 0,
and only the biggest component was used when calculating $C(e)$ and $L(e)$ in those cases.
In Fig.\ \ref{smallworld}, we show $C(e)$ and $L(e)$
[$k(e)\simeq9+10^3e$ for $e\ll1$],
and we see some resemblance with the classic models of SW networks\cite{Watts:1998,Newman:1999}.
Here the set of $e$ can be divided into three regions:
(I) $C(e)$ does not change much while $L(e)$ decreases quickly (dominated by short-range links with the SW phenomena); 
(II) $C(e)$ starts to decrease and reaches the smallest value;
(III) $C(e)$ starts to increase before reaching 1 at $e=1$ (dominated by long-range links).
Meanwhile, $L(e)$ monotonously decreases to 1.
The overall behavior is the same for cases with different $N$'s.

We then introduce a {\em periodic} state $\phi_i$ ($0\le\phi_i<1$) to each node $i$, and
observe dynamics on the network defined above
($\phi_i$ can be an opinion of an agent $i$ in opinion 
dynamics\cite{Ree:2011}, or
$2\pi\phi_i$ can be a phase of an oscillator $i$ in synchronization 
problems\cite{Hong:2002,Heylen:2006,Arenas:2008,DiazGuilera:2009}).
For changes of $\phi_i$'s, we use the rule of repeated averaging as below.
At time $t$, all nodes update their values in parallel using
\begin{equation}
	\phi_i(t+1) = \phi_i(t)+\frac{\sigma}{k_i(t)+1}\sum_j K_{ij}(t)\Delta_{ji}(t)\ \mbox{(mod 1)}, 
	\label{phi_change}
\end{equation}
where $\sigma$ is a coupling constant ($0<\sigma\le1$),
$k_i(t)$ is the degree of node $i$,
and $\Delta_{ji}(t)$ represents the difference between $\phi_j$ and $\phi_i$ at time $t$.
We introduce a function to represent the difference between two $\phi$ values,
\begin{equation}
	\Delta(\phi_j,\phi_i)\equiv\left\{
		\begin{array}{ll}
			\phi_j-\phi_i+1 & (\mbox{if}\ \phi_j-\phi_i\le-0.5),\\
			\phi_j-\phi_i-1 & (\mbox{if}\ \phi_j-\phi_i >0.5),\\
			\phi_j-\phi_i    & (\mbox{otherwise}), 
		\end{array}\right.
	\label{periodic}
\end{equation}
and $\Delta_{ji}\equiv\Delta(\phi_j,\phi_i)$.
Note that Eq.\ (\ref{phi_change}) is the same as the one given in Ref.\ \onlinecite{Ree:2011} 
with different notations, except that there is no threshold for $\Delta_{ji}$, and that, 
if we exchange this $\Delta$-function with the sine function, it basically becomes
the Kuramoto model\cite{Kuramoto:1984,*Acebron:2005}, used in many synchronization problems.

In a special case with $e=0$, it was already found that {\em periodic} metastable states
(along the $x$ direction) can spontaneously emerge from random initial conditions especially when
$X\ll Y$, due to periodicities of both the state variable and the lattice structure\cite{Ree:2011}.
We call them ``period-$n$" states, and they have finite mean lifetimes  $\tau^{(p)}_n$.
When $e\ll1$, $\tau^{(p)}_n$ for low $n$ is too long to obtain numerically, but
we will see that, as $e$ increases, $\tau^{(p)}_n$ decreases 
because long-range links make periodic metastable states
decay faster and eventually disappear.

As $e$ increases further, however, another type of metastable states starts to appear, 
and they are {\em grouping} metastable states,
where nodes are divided into $m$ separate groups of different $\phi$'s, independent of their positions on the lattice.
For these states, numbers of nodes in groups and $\phi$-distances between groups should satisfy 
certain conditions: for example, three groups with similar sizes at $\phi\simeq0, 1/3$, and 2/3, 
respectively, or \Tilde500 nodes at $\phi\simeq0.5$, and 
two other groups of \Tilde250 nodes at $\phi\simeq0.1$ and 0.9, respectively. 
We call them ``$m$-group" states, while $m$ is an odd number\footnote{
Even-numbered groups cannot be stable because of the dynamical rule given 
in Eq.\ (\ref{periodic}), where there
exists uncertainty at $\Delta_{ji}=0.5$.}.
Their mean lifetimes increase as $e$ increases, and as $e$ approaches 1, $m$-group states
with higher $m$'s become more stable.
Meanwhile, {\em converged} states, where state variables of all nodes converge to 
one value, are absorbing states in every case  ({\em consensus} in opinion
dynamics, or {\em synchronized} states in synchronization problems).
If either the state variable or the lattice is not periodic, there is no metastable
state, and the system reaches
a converged state with repeated-averaging mechanisms, while $e$ and $\sigma$ 
only determine the convergence time.
Converged states can be also called as either period-0 or 1-group states.

Here we numerically study period-$n$ states in detail, as well as 3-group states.
Usually, period-$n$ states decay into period-$(n-1)$ states, and later into period-$(n-2)$ states, until they
eventually reach converged states---this is a decay chain.
To investigate this decay process closely, one needs a way to find what period a given state is in. 
We introduce $\nu$ for this purpose.
First, we divide the lattice into $N_r$ equal-sized regions along the $x$ direction.
(We assume $X$ is a multiple of $N_r$, and use $N_r=50$ here.)
Second, for each region $r$ ($0\le r<N_r$), we find the {\em average} of state variables, $\bar\phi_r$, 
of nodes inside that region.
Since the state variable is periodic, care has to be taken; for example, the average of 0.1 and 0.9
should be 0 instead of 0.5 [$\Delta$-function in Eq.\ (\ref{periodic}) should be used].
It is only meaningful when the range of all values in a region is less than a certain value like 0.5, 
which is true in most cases when $e\ll1$, because of the repeated averaging between neighbors.
Once we find all $\bar\phi_r$'s for all $r$, we can finally define $\nu$ as
\begin{equation}
	\nu\equiv \left|\sum^{N_r-1}_{r=0} \Delta(\bar\phi_{r+1},\bar\phi_r)\right|,
	\label{nu}
\end{equation}
where $\bar\phi_{N_r}=\bar\phi_0$.
For period-$n$ states, $\nu$ is close to $n$, and
we can observe behavior of a state by observing the time evolution of $\nu$.

\begin{figure}
\includegraphics[scale=0.58]{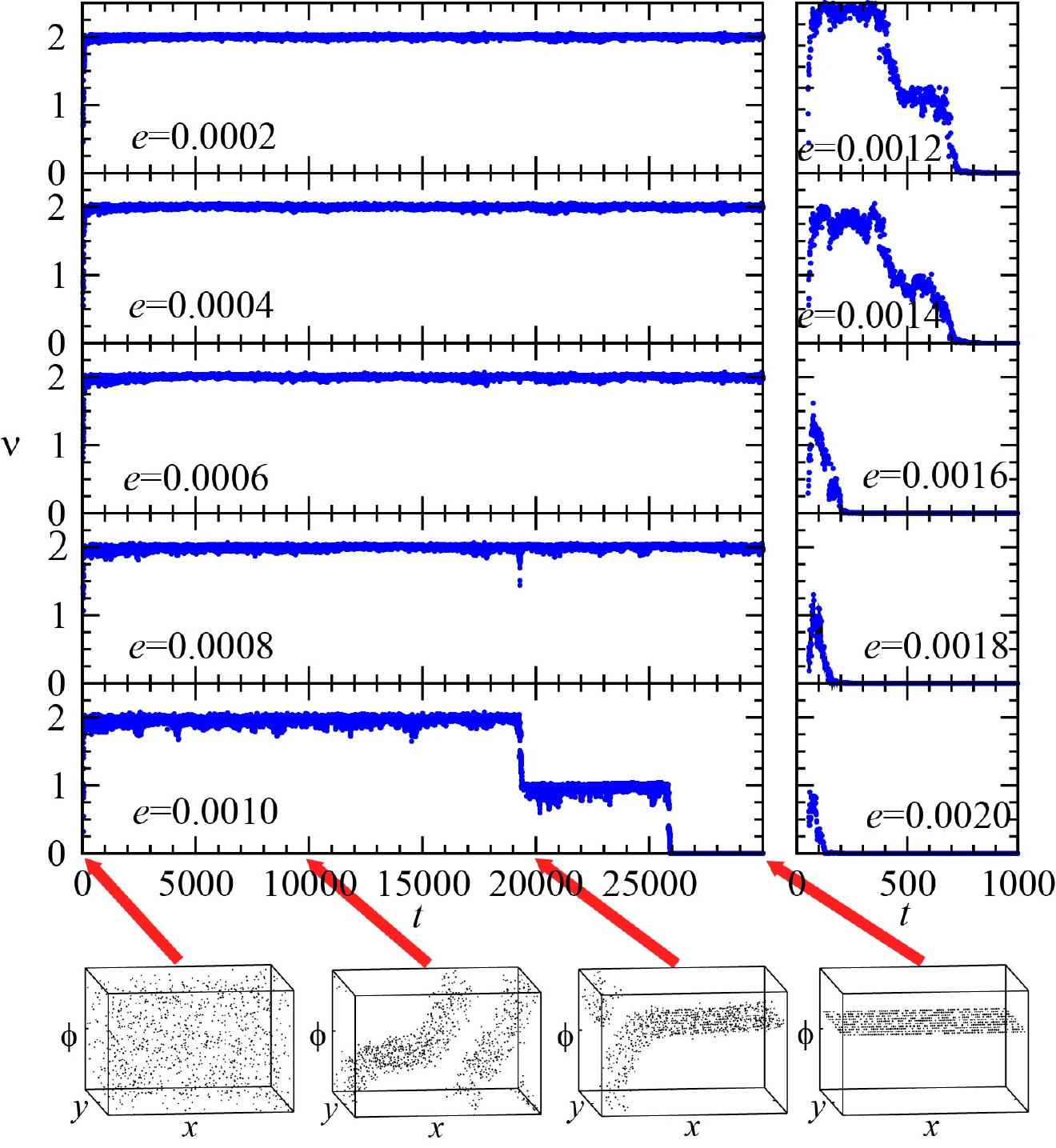}
\caption{\label{behavior}
(color online).
Time evolution of $\nu$ for a state as $e$ increases from $0.0002$ to $0.002$  when $\sigma=0.1$. 
Bottom: Scatter plots for $e=0.001$ at
$t=0$ (random initial condition), $10\ 000$ (period 2), $20\ 000$ (period 1), and $30 \ 000$ (period 0).
	}
\end{figure}

In Fig.\ \ref{behavior},
we observe how a random state evolves for different $e$'s when $\sigma=0.1$.
When $e<0.001$, the state quickly becomes a period-2 state, and 
it does not decay before $t=30\ 000$. 
But when $e=0.001$, we can observe the period-2 state decays into a period-1 state, and then a period-0 state.
Note that lifetimes found here are only for this run, and {\em mean} lifetimes for $e=0.001$ are
shown in Fig.\ 5.
As $e$ increases further, period-$n$ states tend to decay faster, and for higher $e$'s, period-$n$
states are reached only for a short time or bypassed (when $e=0.002$, the state tries to reach 
a period-1 state, but fails before becoming a period-0 state).

\begin{figure}
\includegraphics[scale=0.4]{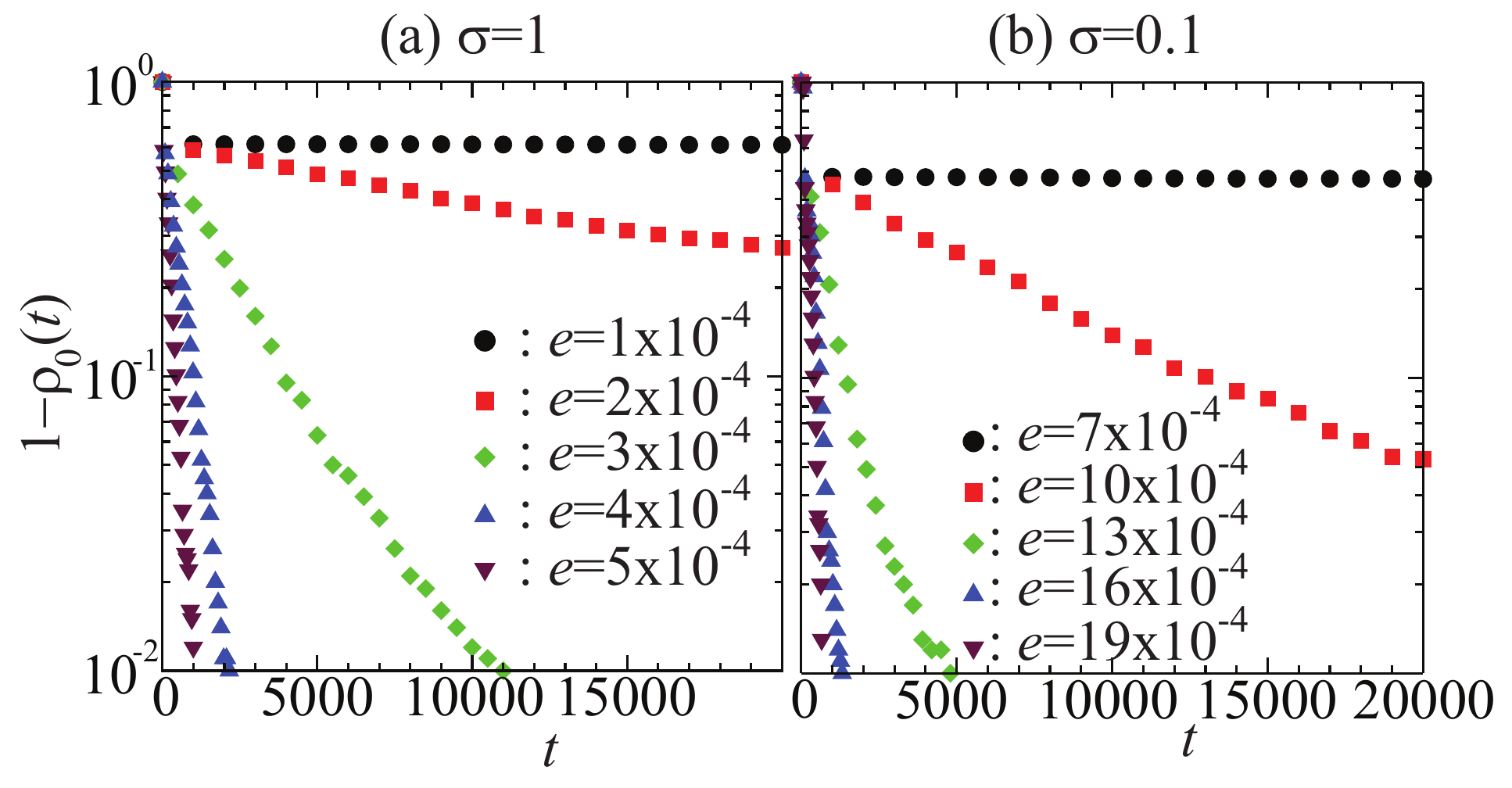}
\caption{\label{randominit}
(color online).
Observing $\rho_0(t)$, the ratio of the number of converged states at $t$ to $10^3$ runs (each with
different random initial condition),
for several $e$ values when (a) $\sigma=1$, and (b) $\sigma=0.1$.
	}
\end{figure}

When $\sigma$ is not too small, the state quickly reaches any period-$n$ state 
or a converged state given an initial condition, and 
distributions of these states depend on all parameters\cite{Ree:2011}.
When mean lifetimes are short enough to be measured,
these distributions visibly change as time evolves, and,
if we use the term $\rho_n(t)$ to represent the ratios of period-$n$ states to all runs,
we can predict that all period-$n$ states decay and $\rho_0(t)$ eventually
converges to $1$ in a reasonable time.
In Fig.\ \ref{randominit},
we use $10^3$ random initial conditions (one run each), and observe
$\rho_0(t)$ as $t$ evolves when $\sigma=1$ and 0.1.
We see similar behavior for both $\sigma$ values. 
Even though decays from $n$ to $n-1$ are all exponential, curves don't have to be straight here,
because period-1 and period-2 states coexist initially and period-2 states go through period-1 states to get to period-0 states. 

\begin{figure}
\includegraphics[scale=0.49]{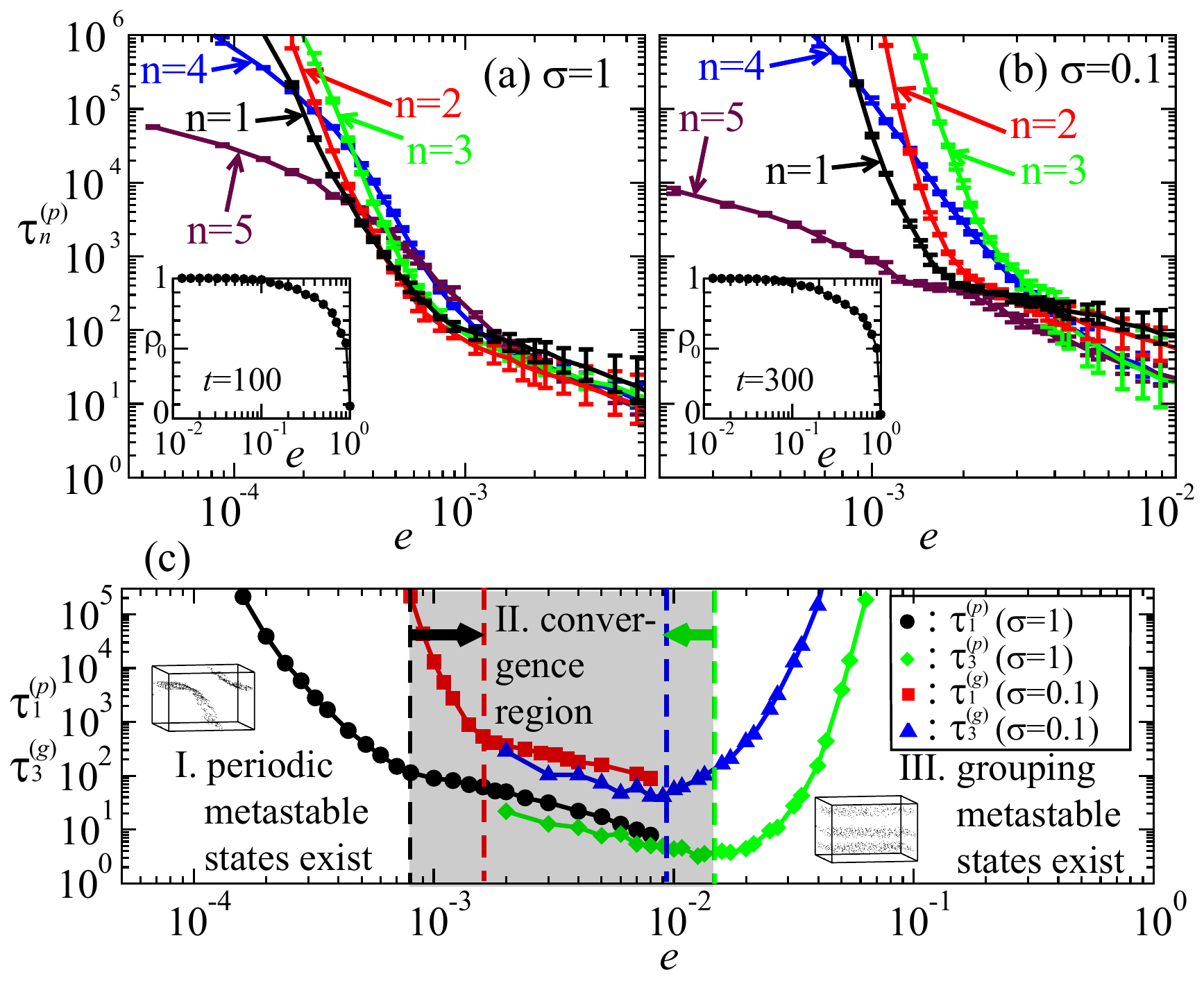}
\caption{\label{decayrates}
(color online).
Behavior of $\tau^{(p)}_n$ ($1\le n \le5$) and $\tau^{(g)}_3$ as $e$ varies.
(a) $\tau^{(p)}_n$ when $\sigma=1$, 
(b) $\tau^{(p)}_n$ when $\sigma=0.1$. 
For each parameter set, we used three different realizations of period-$n$ states with  
$10^3$ runs each.
Bars represent minimum and maximum values.
(Insets: From $10^3$ random initial conditions, 
grouping metastable states appear spontaneously as $e$ gets close to 1.
In both cases, $\rho_0$ is less than 1 when $e>0.1$.)
(c) $\tau^{(p)}_1$ and $\tau^{(g)}_3$ for $0< e\le1$ when $\sigma=0$ and 0.1.
There are three regions of $e$, and the middle region (shaded for the case of $\sigma=1$), 
in which metastable states cannot be sustained, shrinks as $\sigma$ decreases.
	}
\end{figure}

If we specifically choose an initial condition as a period-$n$ state and observe
the behavior of $\rho_{n-1}(t)$ with $10^3$ runs, then curves are straight,
and we can find $\tau^{(p)}_n$ by finding slopes using the method of
linear least squares. 
In Fig.\ \ref{decayrates},
we find $\tau^{(p)}_n$ ($1\le n\le5$) and 
$\tau^{(g)}_3$, the mean lifetime of states with three equal-size groups,
as $e$ is varied when $\sigma=1$ and 0.1.
We found several interesting properties for $\tau^{(p)}_n$ 
in Figs.\ \ref{decayrates}(a) and \ref{decayrates}(b).
For period-$n$ states for low $n$,
there exists a critical $e$ value, $e^{(p)}_n$, 
after which period-$n$ states cannot be sustained---for example,
$e^{(p)}_1\simeq 0.002$ when $\sigma=0.1$ (see Fig.\ \ref{behavior}).
When $e>e^{(p)}_n$, period-$n$ states cannot be sustained for even a short period of time, while
the variability of $\tau^{(p)}_n$ with different realizations becomes greater.
When $e<e^{(p)}_n$, $\tau^{(p)}_n$ does not depend much on initial conditions
as long as they are period-$n$ states.
Higher-period states can be more stable than lower-period ones for certain $e$'s
($\tau^{(p)}_3>\tau^{(p)}_2>\tau^{(p)}_4>\tau^{(p)}_1>\tau^{(p)}_5$ for $\sigma=0.1$ and $e=0.001$), 
but we can predict that the order should become normal
as $e$ approaches 0.
On the other hand, when $e$ gets close to 1, grouping metastable states can exist. 
Insets prove their existence: when $e>0.1$, $\rho_0$ from $10^3$ runs (each 
with a different random initial condition)
can stay less than 1 for a considerable amount of time.
The closer $e$ is to 1, the longer the mean lifetimes of these metastable states become.

In Fig.\ \ref{decayrates}(c), we observe $\tau^{(p)}_1$ and $\tau^{(g)}_3$ for
the whole range of $e$ when $\sigma=0$ and 0.1.
There also exists a critical value $e^{(g)}_3$ for 3-group states.
When $e>e^{(g)}_3$, 3-group states can exist and $\tau^{(g)}_3$ increases with $e$. 
When $e<e^{(g)}_3$, 3-group states cannot be sustained, 
but $\tau^{(g)}_3$ tend to coincide with $\tau^{(p)}_1$, increasing again
as $e$ decreases, because 3-group states try to form period-1 state
before converging to 1-group states without success.
Based on results for $\sigma=1$, we divide $e$ into three regions,
which match well with those found in Fig.\ \ref{smallworld}(a):
(I) $0<e<e^{(p)}_1$, where periodic metastable states can exist 
(dominated by short-range links);
(II) $e^{(p)}_1<e<e^{(g)}_3$, where only converged states are sustained;
(III) $e^{(g)}_3<e\le1$, where grouping metastable states can exist 
(dominated by long-range links).
It is also interesting that the range of $e$, where small-world phenomena 
are observed,
coincides with the range where periodic metastable states have measurable mean lifetimes.
We also observe results for $\sigma=0.1$, and they show similar behavior with 
longer lifetimes and smaller convergence regions.
If $\sigma$ is much smaller, the system can behave in an unexpected manner;
for example,  even when $e=0$, grouping metastable states can appear
(not shown here; to be discussed in future work).

In summary, we studied dynamics of random-walking nodes
with periodic states 
in a periodic lattice, varying the number of long-range links, $e$, and
observed two types of metastable states in the system
at low- and high-$e$ limits, respectively.
Metastable states in our model 
can be interpreted as dynamic, yet sustainable, diversities abundant in our 
societies\cite{Mas:2010},
and our result implies that ever-increasing connections of our modern societies can make existing diversities
disappear, but if we are connected too well, another type of states of diversities can appear.

Metastable states can exist due to periodicities of both the node state and the space,
and we can modify or extend this simple model without changing the overall behavior to suit realistic needs.
One coupling constant $\sigma$ is given to both short- and long-range interactions here,
but we can have two coupling constants\cite{Heylen:2006}, or it can be heterogeneous for all nodes.
We can add noise in Eq.\ (\ref{phi_change}), because noise can play an important 
role in clustering\cite{Mas:2010}.
In addition, long-range links can be rewired based on node states, 
making the model truly coevolving.

The author would like to thank L.\ E.\ Reichl for useful discussions, and 
Kongju National University for financial support.

\bibliography{paper}

\begin{thebibliography}{25}%
\makeatletter
\providecommand \@ifxundefined [1]{%
 \@ifx{#1\undefined}
}%
\providecommand \@ifnum [1]{%
 \ifnum #1\expandafter \@firstoftwo
 \else \expandafter \@secondoftwo
 \fi
}%
\providecommand \@ifx [1]{%
 \ifx #1\expandafter \@firstoftwo
 \else \expandafter \@secondoftwo
 \fi
}%
\providecommand \natexlab [1]{#1}%
\providecommand \enquote  [1]{``#1''}%
\providecommand \bibnamefont  [1]{#1}%
\providecommand \bibfnamefont [1]{#1}%
\providecommand \citenamefont [1]{#1}%
\providecommand \href@noop [0]{\@secondoftwo}%
\providecommand \href [0]{\begingroup \@sanitize@url \@href}%
\providecommand \@href[1]{\@@startlink{#1}\@@href}%
\providecommand \@@href[1]{\endgroup#1\@@endlink}%
\providecommand \@sanitize@url [0]{\catcode `\\12\catcode `\$12\catcode
  `\&12\catcode `\#12\catcode `\^12\catcode `\_12\catcode `\%12\relax}%
\providecommand \@@startlink[1]{}%
\providecommand \@@endlink[0]{}%
\providecommand \url  [0]{\begingroup\@sanitize@url \@url }%
\providecommand \@url [1]{\endgroup\@href {#1}{\urlprefix }}%
\providecommand \urlprefix  [0]{URL }%
\providecommand \Eprint [0]{\href }%
\@ifxundefined \urlstyle {%
  \providecommand \doi  [0]{\begingroup \@sanitize@url \@doi}%
  \providecommand \@doi [1]{\endgroup \@@startlink {\doibase
  #1}doi:\discretionary {}{}{}#1\@@endlink }%
}{%
  \providecommand \doi  [0]{doi:\discretionary{}{}{}\begingroup
  \urlstyle{rm}\Url }%
}%
\providecommand \doibase [0]{http://dx.doi.org/}%
\providecommand \Doi [0]{\begingroup \@sanitize@url \@Doi }%
\providecommand \@Doi  [1]{\endgroup\@@startlink{\doibase#1}\@@Doi}%
\providecommand \@@Doi [1]{#1\@@endlink}%
\providecommand \selectlanguage [0]{\@gobble}%
\providecommand \bibinfo  [0]{\@secondoftwo}%
\providecommand \bibfield  [0]{\@secondoftwo}%
\providecommand \translation [1]{[#1]}%
\providecommand \BibitemOpen [0]{}%
\providecommand \bibitemStop [0]{}%
\providecommand \bibitemNoStop [0]{.\EOS\space}%
\providecommand \EOS [0]{\spacefactor3000\relax}%
\providecommand \BibitemShut  [1]{\csname bibitem#1\endcsname}%
\bibitem [{\citenamefont {Albert}\ and\ \citenamefont
  {Barab{\'a}si}(2002)}]{Albert:2002}%
  \BibitemOpen
  \bibfield  {author} {\bibinfo {author} {\bibfnamefont {R.}~\bibnamefont
  {Albert}}\ and\ \bibinfo {author} {\bibfnamefont {A.-L.}\ \bibnamefont
  {Barab{\'a}si}},\ }\href@noop {} {\bibfield  {journal} {\bibinfo  {journal}
  {Rev.\ Mod.\ Phys.},\ }\textbf {\bibinfo {volume} {74}},\ \bibinfo {pages}
  {47} (\bibinfo {year} {2002})}\BibitemShut {NoStop}%
\bibitem [{\citenamefont {Boccaletti}\ \emph {et~al.}(2006)\citenamefont
  {Boccaletti}, \citenamefont {Latora}, \citenamefont {Moreno}, \citenamefont
  {Chavez},\ and\ \citenamefont {Hwang}}]{Boccaletti:2006}%
  \BibitemOpen
  \bibfield  {author} {\bibinfo {author} {\bibfnamefont {S.}~\bibnamefont
  {Boccaletti}}, \bibinfo {author} {\bibfnamefont {V.}~\bibnamefont {Latora}},
  \bibinfo {author} {\bibfnamefont {Y.}~\bibnamefont {Moreno}}, \bibinfo
  {author} {\bibfnamefont {M.}~\bibnamefont {Chavez}}, \ and\ \bibinfo {author}
  {\bibfnamefont {D.-Y.}\ \bibnamefont {Hwang}},\ }\href@noop {} {\bibfield
  {journal} {\bibinfo  {journal} {Phys.\ Rep.},\ }\textbf {\bibinfo {volume}
  {424}},\ \bibinfo {pages} {175} (\bibinfo {year} {2006})}\BibitemShut
  {NoStop}%
\bibitem [{\citenamefont {Dorogovtsev}\ \emph {et~al.}(2008)\citenamefont
  {Dorogovtsev}, \citenamefont {Goltsev},\ and\ \citenamefont
  {Mendes}}]{Dorogovtsev:2008}%
  \BibitemOpen
  \bibfield  {author} {\bibinfo {author} {\bibfnamefont {S.~N.}\ \bibnamefont
  {Dorogovtsev}}, \bibinfo {author} {\bibfnamefont {A.~V.}\ \bibnamefont
  {Goltsev}}, \ and\ \bibinfo {author} {\bibfnamefont {J.~F.~F.}\ \bibnamefont
  {Mendes}},\ }\href@noop {} {\bibfield  {journal} {\bibinfo  {journal} {Rev.\
  Mod.\ Phys.},\ }\textbf {\bibinfo {volume} {80}},\ \bibinfo {pages} {1275}
  (\bibinfo {year} {2008})}\BibitemShut {NoStop}%
\bibitem [{\citenamefont {Watts}\ and\ \citenamefont
  {Strogatz}(1998)}]{Watts:1998}%
  \BibitemOpen
  \bibfield  {author} {\bibinfo {author} {\bibfnamefont {D.~J.}\ \bibnamefont
  {Watts}}\ and\ \bibinfo {author} {\bibfnamefont {S.~H.}\ \bibnamefont
  {Strogatz}},\ }\href@noop {} {\bibfield  {journal} {\bibinfo  {journal}
  {Nature},\ }\textbf {\bibinfo {volume} {393}},\ \bibinfo {pages} {440}
  (\bibinfo {year} {1998})}\BibitemShut {NoStop}%
\bibitem [{\citenamefont {Newman}\ and\ \citenamefont
  {Watts}(1999)}]{Newman:1999}%
  \BibitemOpen
  \bibfield  {author} {\bibinfo {author} {\bibfnamefont {M.~E.~J.}\
  \bibnamefont {Newman}}\ and\ \bibinfo {author} {\bibfnamefont {D.~J.}\
  \bibnamefont {Watts}},\ }\href@noop {} {\bibfield  {journal} {\bibinfo
  {journal} {Phys.\ Rev.\ E},\ }\textbf {\bibinfo {volume} {60}},\ \bibinfo
  {pages} {7332} (\bibinfo {year} {1999})}\BibitemShut {NoStop}%
\bibitem [{\citenamefont {Barab{\'a}si}\ and\ \citenamefont
  {Albert}(1999)}]{Barabasi:1999}%
  \BibitemOpen
  \bibfield  {author} {\bibinfo {author} {\bibfnamefont {A.-L.}\ \bibnamefont
  {Barab{\'a}si}}\ and\ \bibinfo {author} {\bibfnamefont {R.}~\bibnamefont
  {Albert}},\ }\href@noop {} {\bibfield  {journal} {\bibinfo  {journal}
  {Science},\ }\textbf {\bibinfo {volume} {286}},\ \bibinfo {pages} {509}
  (\bibinfo {year} {1999})}\BibitemShut {NoStop}%
\bibitem [{\citenamefont {Barahona}\ and\ \citenamefont
  {Pecora}(2002)}]{Barahona:2002}%
  \BibitemOpen
  \bibfield  {author} {\bibinfo {author} {\bibfnamefont {M.}~\bibnamefont
  {Barahona}}\ and\ \bibinfo {author} {\bibfnamefont {L.~M.}\ \bibnamefont
  {Pecora}},\ }\href@noop {} {\bibfield  {journal} {\bibinfo  {journal} {Phys.\
  Rev.\ Lett.},\ }\textbf {\bibinfo {volume} {89}},\ \bibinfo {pages} {054101}
  (\bibinfo {year} {2002})}\BibitemShut {NoStop}%
\bibitem [{\citenamefont {Laguna}\ \emph {et~al.}(2003)\citenamefont {Laguna},
  \citenamefont {Abramson},\ and\ \citenamefont {Zanette}}]{Laguna:2003}%
  \BibitemOpen
  \bibfield  {author} {\bibinfo {author} {\bibfnamefont {M.~F.}\ \bibnamefont
  {Laguna}}, \bibinfo {author} {\bibfnamefont {G.}~\bibnamefont {Abramson}}, \
  and\ \bibinfo {author} {\bibfnamefont {D.~H.}\ \bibnamefont {Zanette}},\
  }\href@noop {} {\bibfield  {journal} {\bibinfo  {journal} {Physica A},\
  }\textbf {\bibinfo {volume} {329}},\ \bibinfo {pages} {459} (\bibinfo {year}
  {2003})}\BibitemShut {NoStop}%
\bibitem [{\citenamefont {Suchecki}\ \emph {et~al.}(2005)\citenamefont
  {Suchecki}, \citenamefont {Egu{\'\i}luz},\ and\ \citenamefont
  {San~Miguel}}]{Suchecki:2005}%
  \BibitemOpen
  \bibfield  {author} {\bibinfo {author} {\bibfnamefont {K.}~\bibnamefont
  {Suchecki}}, \bibinfo {author} {\bibfnamefont {V.~M.}\ \bibnamefont
  {Egu{\'\i}luz}}, \ and\ \bibinfo {author} {\bibfnamefont {M.}~\bibnamefont
  {San~Miguel}},\ }\href@noop {} {\bibfield  {journal} {\bibinfo  {journal}
  {Phys.\ Rev.\ E},\ }\textbf {\bibinfo {volume} {72}},\ \bibinfo {pages}
  {036132} (\bibinfo {year} {2005})}\BibitemShut {NoStop}%
\bibitem [{\citenamefont {Toivonen}\ \emph {et~al.}(2009)\citenamefont
  {Toivonen}, \citenamefont {Castell{\'o}}, \citenamefont {Egu{\'\i}luz},
  \citenamefont {Saram{\"a}ki}, \citenamefont {Kaski},\ and\ \citenamefont
  {San~Miguel}}]{Toivonen:2009}%
  \BibitemOpen
  \bibfield  {author} {\bibinfo {author} {\bibfnamefont {R.}~\bibnamefont
  {Toivonen}}, \bibinfo {author} {\bibfnamefont {X.}~\bibnamefont
  {Castell{\'o}}}, \bibinfo {author} {\bibfnamefont {V.~M.}\ \bibnamefont
  {Egu{\'\i}luz}}, \bibinfo {author} {\bibfnamefont {J.}~\bibnamefont
  {Saram{\"a}ki}}, \bibinfo {author} {\bibfnamefont {K.}~\bibnamefont {Kaski}},
  \ and\ \bibinfo {author} {\bibfnamefont {M.}~\bibnamefont {San~Miguel}},\
  }\href@noop {} {\bibfield  {journal} {\bibinfo  {journal} {Phys Rev E},\
  }\textbf {\bibinfo {volume} {79}},\ \bibinfo {pages} {016109} (\bibinfo
  {year} {2009})}\BibitemShut {NoStop}%
\bibitem [{\citenamefont {Hong}\ \emph {et~al.}(2002)\citenamefont {Hong},
  \citenamefont {Choi},\ and\ \citenamefont {Kim}}]{Hong:2002}%
  \BibitemOpen
  \bibfield  {author} {\bibinfo {author} {\bibfnamefont {H.}~\bibnamefont
  {Hong}}, \bibinfo {author} {\bibfnamefont {M.~Y.}\ \bibnamefont {Choi}}, \
  and\ \bibinfo {author} {\bibfnamefont {B.~J.}\ \bibnamefont {Kim}},\
  }\href@noop {} {\bibfield  {journal} {\bibinfo  {journal} {Phys.\ Rev.\ E},\
  }\textbf {\bibinfo {volume} {65}},\ \bibinfo {pages} {026139} (\bibinfo
  {year} {2002})}\BibitemShut {NoStop}%
\bibitem [{\citenamefont {Heylen}\ \emph {et~al.}(2006)\citenamefont {Heylen},
  \citenamefont {Skantzos}, \citenamefont {Busquets~Blanco},\ and\
  \citenamefont {Boll{\'e}}}]{Heylen:2006}%
  \BibitemOpen
  \bibfield  {author} {\bibinfo {author} {\bibfnamefont {R.}~\bibnamefont
  {Heylen}}, \bibinfo {author} {\bibfnamefont {N.~S.}\ \bibnamefont
  {Skantzos}}, \bibinfo {author} {\bibfnamefont {J.}~\bibnamefont
  {Busquets~Blanco}}, \ and\ \bibinfo {author} {\bibfnamefont {D.}~\bibnamefont
  {Boll{\'e}}},\ }\href@noop {} {\bibfield  {journal} {\bibinfo  {journal}
  {Phys.\ Rev.\ E},\ }\textbf {\bibinfo {volume} {73}},\ \bibinfo {pages}
  {016138} (\bibinfo {year} {2006})}\BibitemShut {NoStop}%
\bibitem [{\citenamefont {Holme}\ and\ \citenamefont
  {Newman}(2006)}]{Holme:2006}%
  \BibitemOpen
  \bibfield  {author} {\bibinfo {author} {\bibfnamefont {P.}~\bibnamefont
  {Holme}}\ and\ \bibinfo {author} {\bibfnamefont {M.~E.~J.}\ \bibnamefont
  {Newman}},\ }\href@noop {} {\bibfield  {journal} {\bibinfo  {journal} {Phys.\
  Rev.\ E},\ }\textbf {\bibinfo {volume} {74}},\ \bibinfo {pages} {056108}
  (\bibinfo {year} {2006})}\BibitemShut {NoStop}%
\bibitem [{\citenamefont {Gonz{\'a}lez}\ \emph {et~al.}(2006)\citenamefont
  {Gonz{\'a}lez}, \citenamefont {Lind},\ and\ \citenamefont
  {Herrmann}}]{Gonzalez:2006}%
  \BibitemOpen
  \bibfield  {author} {\bibinfo {author} {\bibfnamefont {M.~C.}\ \bibnamefont
  {Gonz{\'a}lez}}, \bibinfo {author} {\bibfnamefont {P.~G.}\ \bibnamefont
  {Lind}}, \ and\ \bibinfo {author} {\bibfnamefont {H.~J.}\ \bibnamefont
  {Herrmann}},\ }\href@noop {} {\bibfield  {journal} {\bibinfo  {journal}
  {Phys.\ Rev.\ Lett.},\ }\textbf {\bibinfo {volume} {96}},\ \bibinfo {pages}
  {088702} (\bibinfo {year} {2006})}\BibitemShut {NoStop}%
\bibitem [{\citenamefont {Benczik}\ \emph {et~al.}(2009)\citenamefont
  {Benczik}, \citenamefont {Benczik}, \citenamefont {Schmittmann},\ and\
  \citenamefont {Zia}}]{Benczik:2009}%
  \BibitemOpen
  \bibfield  {author} {\bibinfo {author} {\bibfnamefont {I.~J.}\ \bibnamefont
  {Benczik}}, \bibinfo {author} {\bibfnamefont {S.~Z.}\ \bibnamefont
  {Benczik}}, \bibinfo {author} {\bibfnamefont {B.}~\bibnamefont
  {Schmittmann}}, \ and\ \bibinfo {author} {\bibfnamefont {R.~K.~P.}\
  \bibnamefont {Zia}},\ }\href@noop {} {\bibfield  {journal} {\bibinfo
  {journal} {Phys.\ Rev.\ E},\ }\textbf {\bibinfo {volume} {79}},\ \bibinfo
  {pages} {046104} (\bibinfo {year} {2009})}\BibitemShut {NoStop}%
\bibitem [{\citenamefont {Manrubia}\ \emph {et~al.}(2001)\citenamefont
  {Manrubia}, \citenamefont {Delgado},\ and\ \citenamefont
  {Luque}}]{Manrubia:2001}%
  \BibitemOpen
  \bibfield  {author} {\bibinfo {author} {\bibfnamefont {S.~C.}\ \bibnamefont
  {Manrubia}}, \bibinfo {author} {\bibfnamefont {J.}~\bibnamefont {Delgado}}, \
  and\ \bibinfo {author} {\bibfnamefont {B.}~\bibnamefont {Luque}},\
  }\href@noop {} {\bibfield  {journal} {\bibinfo  {journal} {Europhys.\ Lett},\
  }\textbf {\bibinfo {volume} {53}},\ \bibinfo {pages} {693} (\bibinfo {year}
  {2001})}\BibitemShut {NoStop}%
\bibitem [{\citenamefont {Fujiwara}\ \emph {et~al.}(2011)\citenamefont
  {Fujiwara}, \citenamefont {Kurths},\ and\ \citenamefont
  {D{\'\i}az-Guilera}}]{Fujiwara:2011}%
  \BibitemOpen
  \bibfield  {author} {\bibinfo {author} {\bibfnamefont {N.}~\bibnamefont
  {Fujiwara}}, \bibinfo {author} {\bibfnamefont {J.}~\bibnamefont {Kurths}}, \
  and\ \bibinfo {author} {\bibfnamefont {A.}~\bibnamefont
  {D{\'\i}az-Guilera}},\ }\href@noop {} {\bibfield  {journal} {\bibinfo
  {journal} {Phys.\ Rev.\ E},\ }\textbf {\bibinfo {volume} {83}},\ \bibinfo
  {pages} {025101} (\bibinfo {year} {2011})}\BibitemShut {NoStop}%
\bibitem [{\citenamefont {Ree}(2011)}]{Ree:2011}%
  \BibitemOpen
  \bibfield  {author} {\bibinfo {author} {\bibfnamefont {S.}~\bibnamefont
  {Ree}},\ }\href@noop {} {\bibfield  {journal} {\bibinfo  {journal} {Phys.\
  Rev.\ E},\ }\textbf {\bibinfo {volume} {83}},\ \bibinfo {pages} {056110}
  (\bibinfo {year} {2011})}\BibitemShut {NoStop}%
\bibitem [{Note1()}]{Note1}%
  \BibitemOpen
  \bibinfo {note} {Because the motion of nodes is diffusive, the density of
  nodes on the lattice converges to the uniform distribution regardless of the
  initial positions of nodes.}\BibitemShut {Stop}%
\bibitem [{\citenamefont {Arenas}\ \emph {et~al.}(2008)\citenamefont {Arenas},
  \citenamefont {D{\'\i}az-Guilera}, \citenamefont {Kurths}, \citenamefont
  {Moreno},\ and\ \citenamefont {Zhou}}]{Arenas:2008}%
  \BibitemOpen
  \bibfield  {author} {\bibinfo {author} {\bibfnamefont {A.}~\bibnamefont
  {Arenas}}, \bibinfo {author} {\bibfnamefont {A.}~\bibnamefont
  {D{\'\i}az-Guilera}}, \bibinfo {author} {\bibfnamefont {J.}~\bibnamefont
  {Kurths}}, \bibinfo {author} {\bibfnamefont {Y.}~\bibnamefont {Moreno}}, \
  and\ \bibinfo {author} {\bibfnamefont {C.}~\bibnamefont {Zhou}},\ }\href@noop
  {} {\bibfield  {journal} {\bibinfo  {journal} {Phys.\ Rep.},\ }\textbf
  {\bibinfo {volume} {469}},\ \bibinfo {pages} {93} (\bibinfo {year}
  {2008})}\BibitemShut {NoStop}%
\bibitem [{\citenamefont {D{\'\i}az-Guilera}\ \emph {et~al.}(2009)\citenamefont
  {D{\'\i}az-Guilera}, \citenamefont {G{\'o}mez-Garde{\~n}es}, \citenamefont
  {Moreno},\ and\ \citenamefont {Nekovee}}]{DiazGuilera:2009}%
  \BibitemOpen
  \bibfield  {author} {\bibinfo {author} {\bibfnamefont {A.}~\bibnamefont
  {D{\'\i}az-Guilera}}, \bibinfo {author} {\bibfnamefont {J.}~\bibnamefont
  {G{\'o}mez-Garde{\~n}es}}, \bibinfo {author} {\bibfnamefont {Y.}~\bibnamefont
  {Moreno}}, \ and\ \bibinfo {author} {\bibfnamefont {M.}~\bibnamefont
  {Nekovee}},\ }\href@noop {} {\bibfield  {journal} {\bibinfo  {journal} {Int.\
  J.\ Bifurcat.\ Chaos},\ }\textbf {\bibinfo {volume} {19}},\ \bibinfo {pages}
  {687} (\bibinfo {year} {2009})}\BibitemShut {NoStop}%
\bibitem [{\citenamefont {Kuramoto}(1984)}]{Kuramoto:1984}%
  \BibitemOpen
  \bibfield  {author} {\bibinfo {author} {\bibfnamefont {Y.}~\bibnamefont
  {Kuramoto}},\ }\href@noop {} {\emph {\bibinfo {title} {Chemical Oscillations,
  Waves and Turbulence}}}\ (\bibinfo  {publisher} {Springer},\ \bibinfo
  {address} {New York},\ \bibinfo {year} {1984})\BibitemShut {NoStop}%
\bibitem [{\citenamefont {Acebr{\'o}n}\ \emph {et~al.}(2005)\citenamefont
  {Acebr{\'o}n}, \citenamefont {Bonilla}, \citenamefont {P{\'e}rez~Vicente},
  \citenamefont {Ritort},\ and\ \citenamefont {Spigler}}]{Acebron:2005}%
  \BibitemOpen
  \bibfield  {author} {\bibinfo {author} {\bibfnamefont {J.~A.}\ \bibnamefont
  {Acebr{\'o}n}}, \bibinfo {author} {\bibfnamefont {L.~L.}\ \bibnamefont
  {Bonilla}}, \bibinfo {author} {\bibfnamefont {C.~J.}\ \bibnamefont
  {P{\'e}rez~Vicente}}, \bibinfo {author} {\bibfnamefont {F.}~\bibnamefont
  {Ritort}}, \ and\ \bibinfo {author} {\bibfnamefont {R.}~\bibnamefont
  {Spigler}},\ }\href@noop {} {\bibfield  {journal} {\bibinfo  {journal} {Rev.\
  Mod.\ Phys},\ }\textbf {\bibinfo {volume} {77}},\ \bibinfo {pages} {137}
  (\bibinfo {year} {2005})}\BibitemShut {NoStop}%
\bibitem [{Note2()}]{Note2}%
  \BibitemOpen
  \bibinfo {note} {Even-numbered groups cannot be stable because of the
  dynamical rule given in Eq.\ (\ref {periodic}), where there exists
  uncertainty at $\Delta _{ji}=0.5$.}\BibitemShut {Stop}%
\bibitem [{\citenamefont {M{\"a}s}\ \emph {et~al.}(2010)\citenamefont
  {M{\"a}s}, \citenamefont {Flache},\ and\ \citenamefont {Helbing}}]{Mas:2010}%
  \BibitemOpen
  \bibfield  {author} {\bibinfo {author} {\bibfnamefont {M.}~\bibnamefont
  {M{\"a}s}}, \bibinfo {author} {\bibfnamefont {A.}~\bibnamefont {Flache}}, \
  and\ \bibinfo {author} {\bibfnamefont {D.}~\bibnamefont {Helbing}},\
  }\href@noop {} {\bibfield  {journal} {\bibinfo  {journal} {PLoS Comp.\
  Biol.},\ }\textbf {\bibinfo {volume} {6}},\ \bibinfo {pages} {e1000959}
  (\bibinfo {year} {2010})}\BibitemShut {NoStop}%
\end{thebibliography}%

\end{document}